\documentclass{article}%
\usepackage{amsfonts}
\usepackage{amsmath}
\usepackage{amssymb}
\usepackage{graphicx}%
\providecommand{\U}[1]{\protect\rule{.1in}{.1in}}
\begin{document}

\title{On structure of the polarization operator in a  magnetic field}
\author{V.M. Katkov\\Budker Institute of Nuclear Physics,\\Novosibirsk, 630090, Russia\\e-mail: katkov@inp.nsk.su}
\maketitle

\begin{abstract}
The polarization operator is investigated at arbitrary photon energy in a
constant and homogeneous magnetic field for the strength $H$  less than the
Schwinger critical value. The effective mass of a real photon with a preset
polarization is considered in the quantum energy region as well as in the
quasiclassical one. Obtained in the quantum region expressions  include the
singular terms at the creation threshold of electron and positron on Landau levels.

\end{abstract}

\textbf{1}. In 1971, Adler \cite{[1]} derived the polarization operator of a
photon in a magnetic field using the intrinsic time technique developed by
Schwinger \cite{[2]}. In the energy region lower the pair creation threshold,
the polarization operator in a magnetic field was investigated well enough
(see, for example, the papers \cite{[3]} and the bibliography cited there).
Here, we consider the polarization operator on mass shell ( $k^{2}=0,$the
metric $ab=$ $a^{0}b^{0}-\mathbf{ab}$ is used ) at arbitrary value of the
photon energy $\omega$ in case of a weak magnetic field $H$ (compared with the
critical field $H_{0}=m^{2}/e\ =$ $4,41\cdot10^{13}$ $%
%TCIMACRO{\unit{G}}%
%BeginExpansion
\operatorname{G}%
%EndExpansion
,$ the system of units $\hbar=c=1$ is used).

Our analysis is based on the expression for the polarization operator obtained
in a diagonal form in \cite{[4]} (see Eqs. (3.19), (3.33)). In case of a pure
magnetic field, we have the following expression presented in a covariant form%

\begin{align}
\Pi^{\mu\nu}  &  =-\sum_{i=2,3}\kappa_{i}\beta_{i}^{\mu}\beta_{i}^{\nu
},\ \ \ \beta_{i}\beta_{j}=-\ \delta_{ij},\ \ \ \beta_{i}k=0;\label{1}\\
\beta_{2}^{\mu}  &  =(F^{\ast}k)^{\mu}/\sqrt{-(F^{\ast}k)^{2}},\ \ \ \beta
_{3}^{\mu}=(Fk)^{\mu}/\sqrt{-(F^{\ast}k)^{2}},\ \nonumber\\
\ FF^{\ast}  &  =0,\ \ \ F^{2}=F^{\mu\nu}F_{\mu\nu}=2(H^{2}-E^{2})>0,
\label{01}%
\end{align}
where $F^{\mu\nu}$is the electromagnetic field tensor , $F^{\ast\mu\nu}$ is
the dual tensor, $k^{\mu}$ is the photon momentum, $(Fk)^{\mu}=F^{\mu\nu
}k_{\nu},$%

\begin{equation}
\kappa_{i}=\frac{\alpha}{\pi}m^{2}r%
%TCIMACRO{\tint \limits_{-1}^{1}}%
%BeginExpansion
{\textstyle\int\limits_{-1}^{1}}
%EndExpansion
dv%
%TCIMACRO{\tint \limits_{0}^{\infty-\mathrm{i0}}}%
%BeginExpansion
{\textstyle\int\limits_{0}^{\infty-\mathrm{i0}}}
%EndExpansion
f_{i}(v,x)\exp[\mathrm{i}\psi(v,x)]dx. \label{2}%
\end{equation}
Here%

\begin{align}
f_{2}(v,x)  &  =2\frac{\cos(vx)-\cos x}{\sin^{3}x}-\frac{\cos(vx)}{\sin
x}+v\frac{\cos x\sin(vx)}{\sin^{2}x},\nonumber\\
f_{3}(v,x)  &  =\frac{\cos(vx)}{\sin x}-v\frac{\cos x\sin(vx)}{\sin^{2}%
x}-(1-v^{2})\cot x,\nonumber\\
\psi(v,x)  &  =\frac{1}{\mu}\left\{  2r\frac{\cos x-\cos(vx)}{\sin
x}+[r(1-v^{2})-1]x\right\}  ;\label{3}\\
r  &  =-(F^{\ast}k)^{2}/2m^{2}F^{2},\ \ \ \mu^{2}=F^{2}/2H_{0}^{2}. \label{03}%
\end{align}
The real part of $\kappa_{i}$ determines the refractive index $n_{i}$ of
photon with the polarization $e_{i}=\beta_{i}$:%

\begin{equation}
\ \ \ \ n_{i}=1-\frac{\mathrm{\operatorname{Re}}\kappa_{i}}{2\omega^{2}}.
\label{4}%
\end{equation}
At $r>1,$ the proper value of polarization operator $\kappa_{i}$ includes the
imaginary part, which determines the probability per unit length of pair production:%

\begin{equation}
W_{i}=-\frac{1}{\omega}\mathrm{\operatorname{Im}}\kappa_{i} \label{5}%
\end{equation}
At $r<1,$ the integration counter over $x$ in Eq. (\ref{2}) can be turn to the
lower semiaxis $(x\rightarrow-\mathrm{i}x),$ then the value $\kappa_{i}$
becomes real in the explicit form.

\textbf{2}. We consider now the case of weak field and low energy: $\mu\ll1,$
$1<r\ll1/\mu^{2}.$ Let's remove the integration counter over $x$ in Eq.
(\ref{2}) to the lower semiaxis at the value $x_{0}:$%

\begin{equation}
x_{0}(r)=-\mathrm{i}l(r),\ \ \ \ l(r)=\ln\frac{\sqrt{r}+1}{\sqrt{r}-1}.
\label{6}%
\end{equation}
So we have the following expression for $\kappa_{i}:$%

\begin{equation}
\kappa_{i}=\frac{\alpha}{\pi}m^{2}r\left(  a_{i}+b_{i}\right)  , \label{06}%
\end{equation}
where%

\begin{align}
a_{i}  &  =-\mathrm{i}%
%TCIMACRO{\tint \limits_{-1}^{1}}%
%BeginExpansion
{\textstyle\int\limits_{-1}^{1}}
%EndExpansion
dv%
%TCIMACRO{\tint \limits_{0}^{l(r)}}%
%BeginExpansion
{\textstyle\int\limits_{0}^{l(r)}}
%EndExpansion
dxf_{i}(v,-\mathrm{i}x)\exp[\mathrm{i}\psi(v,-\mathrm{i}x)],\label{7}\\
b_{i}  &  =%
%TCIMACRO{\tint \limits_{-1}^{1}}%
%BeginExpansion
{\textstyle\int\limits_{-1}^{1}}
%EndExpansion
dv%
%TCIMACRO{\tint \limits_{0}^{\infty}}%
%BeginExpansion
{\textstyle\int\limits_{0}^{\infty}}
%EndExpansion
dzf_{i}(v,z+x_{0})\exp[\mathrm{i}\psi(v,z+x_{0})]. \label{07}%
\end{align}
In the integral $a_{i}$ in Eq. (\ref{7}), the small values $x\sim\mu$
contribute. We calculate this integral expanding the entering functions over
$x.$ Taking into account that in the region under consideration the condition
$r\mu^{2}\ll1$ is fulfilled, we keep in the exponent argument the term
$-x/\mu$ only and extend the integration over $x$ to infinity. In the result
of not complicated integration over $v,$ we have:%

\begin{align}
a_{2}  &  =-\frac{16}{45}\mu^{2},\ \ \ a_{3}=-\frac{28}{45}\mu^{2}%
,\ \ \nonumber\\
\ \kappa_{2}^{a}  &  =-\frac{4\alpha m^{2}\kappa^{2}}{45\pi},\ \ \ \kappa
_{3}^{a}=-\frac{7\alpha m^{2}\kappa^{2}}{45\pi},\ \ \ \ \kappa^{2}%
=-\frac{(Fk)^{2}}{H_{0}^{2}m^{2}}. \label{8}%
\end{align}
These asymptotic are well known (see for example \cite{[3]} and the
bibliography cited there).

In the integral $b_{i}$ Eq. (\ref{07}), the small values $v$ contribute.
Expanding entering functions over $v$ and extending the integration over $v$
to infinity, we have%

\begin{align}
b_{i}  &  =\sqrt{\mu\pi}\exp\left(  -\mathrm{i}\frac{\pi}{4}\right)
%TCIMACRO{\dint \limits_{0}^{\infty}}%
%BeginExpansion
{\displaystyle\int\limits_{0}^{\infty}}
%EndExpansion
\frac{dzf_{i}(0,x_{0}+z)}{\sqrt{\chi(x_{0}+z)}}\exp\left[  -\frac{\mathrm{i}%
}{\mu}\varphi(x_{0}+z)\right]  ,\nonumber\\
\varphi(x)  &  =2r\tan\left(  \frac{x}{2}\right)  +(1-r)x,\ \ \ \chi
(x)=rx\left(  1-\frac{x}{\sin x}\right)  . \label{9}%
\end{align}

We consider now the energy region where $r-1\ll1$ when the moving of created
particles is nonrelativistic. In this case%

\begin{align}
\mathrm{i}\varphi(z+x_{0})  &  \simeq\beta(r)+(r-1)(1-e^{-\mathrm{i}%
z}-\mathrm{i}z),\ \ \ \beta(r)=2\sqrt{r}-(r-1)l(r),\nonumber\\
\chi(x)  &  \simeq z+x_{0},\ \ \ f_{2}(0,x)\simeq0,\ \ \ f_{3}(0,x)\simeq
-\mathrm{i.} \label{10}%
\end{align}
In the threshold region, where the particles occupy not very high energy
levels we present Eq. (\ref{9}) for $b_{3}$ in the form%

\begin{align}
b_{3}  &  =-\mathrm{i}\sqrt{\mu\pi}\exp\left(  -\mathrm{i}\frac{\pi}%
{4}\right)  \exp\left[  -\frac{\beta(r)}{\mu}-\gamma\right]
%TCIMACRO{\tint \limits_{0}^{\infty}}%
%BeginExpansion
{\textstyle\int\limits_{0}^{\infty}}
%EndExpansion
\frac{dz}{\sqrt{x_{0}+z}}%
%TCIMACRO{\tsum \limits_{n=0}^{\infty}}%
%BeginExpansion
{\textstyle\sum\limits_{n=0}^{\infty}}
%EndExpansion
\frac{\gamma^{n}}{n!}\exp[\mathrm{i}(\gamma-n)z],\ \ \nonumber\\
\gamma &  =(r-1)/\mu\sim1. \label{11}%
\end{align}
The integral in Eq. (\ref{11}) has a root singularity at whole numbers of
$\gamma.$ For $|\gamma-n|\ \ll x_{0}^{-1}$, $z\sim$ $|\gamma-n|^{-1}>>x_{0},$
we have:%

\begin{align}
b_{3}  &  =-2\mathrm{i}\sqrt{\mu\pi}\exp\left(  -\mathrm{i}\frac{\pi}%
{4}\right)  \exp\left[  -\frac{\beta(r)}{\mu}-\gamma\right]  \frac{\gamma^{n}%
}{n!}%
%TCIMACRO{\tint \limits_{0}^{\infty}}%
%BeginExpansion
{\textstyle\int\limits_{0}^{\infty}}
%EndExpansion
dy\exp[\mathrm{i}(\gamma-n)y^{2}]\nonumber\\
=  &  -\pi\sqrt{\frac{\mu}{|\gamma-n|}}\exp\left[  -\frac{\beta(r)}{\mu
}-\gamma+\mathrm{i}\frac{\pi}{2}\vartheta(\gamma-n)\right]  \frac{n^{n}}{n!},
\label{12}%
\end{align}
where $\vartheta(z)$ is Heaviside function: $\vartheta(z)=1$ for
$z\geqslant0,$ $\vartheta(z)=0$ for $z<0.$ The expression for $\kappa_{3}$
with the accepted accuracy can be rewritten in the following form:%

\begin{equation}
\kappa_{3}^{b}\simeq-\frac{\alpha m^{2}\mu}{\sqrt{|g|}}e^{-\zeta}\frac
{(2\zeta)^{n}}{n!}\exp\left[  \mathrm{i}\frac{\pi}{2}\vartheta(g)\right]
,\ \ g=r-1-n\mu,\ \ \zeta=2r/\mu. \label{13}%
\end{equation}
Far from singularity, the real part of $\kappa_{3}^{b}$ is small compared to
$\kappa_{3}^{a}$ (\ref{8}), but the main term of the imaginary part of the
effective mass is given by Eq. (\ref{13}).

At $\gamma\gg1,$ the small $z$ contributes to the integral in Eq. (\ref{9})$,$ then:%

\begin{align}
\mathrm{i}\varphi(z+x_{0})  &  \simeq\beta(r)+(r-1)z^{2}/2,\ \ \mathrm{i}%
\chi(z+x_{0})\ \simeq\mathrm{i}x_{0}=l(r),\nonumber\\
\kappa_{3}^{b}  &  \simeq-\mathrm{i}\frac{\alpha m^{2}\mu}{\sqrt{2(r-1)l(r)}%
}\exp\left(  -\frac{\beta(r)}{\mu}\right)  . \label{14}%
\end{align}
When the condition $1\lesssim r-1\ll\mu^{-2}$ is fulfilled, in Eq. (\ref{07})
for $b_{i}$ one can carry out the expansion over $v$ and $z$ from the very
outset. As a result we have (see \cite{[5]}, Eq. (B5)):%

\begin{equation}
\kappa_{3}^{b}\simeq-\mathrm{i}\frac{\alpha m^{2}\mu\sqrt{r}}{\sqrt
{\beta(r)(r-1)l(r)}}\exp\left(  \mathrm{-}\frac{\beta(r)}{\mu}\right)
,\ \ \ \kappa_{2}^{b}=\frac{r-1}{2r}\kappa_{3}^{b}. \label{014}%
\end{equation}

\textbf{4}. The region of weak field and high energy ($\mu\ll1,$
$r\gtrsim1/\mu^{2}$) is contained in the region of the standard quasiclassical
approximation (SQA) \cite{[5]}.\ The main contribution to the integral in Eq.
(\ref{2}) is given by small values of $x.$ Expanding the entering functions
Eq. (\ref{3}) over $x$, and carrying out the change of variable $x=\mu t,$ we get:%

\begin{align}
\kappa_{i}  &  =\frac{\alpha m^{2}\kappa^{2}}{24\pi}%
%TCIMACRO{\tint \limits_{0}^{1}}%
%BeginExpansion
{\textstyle\int\limits_{0}^{1}}
%EndExpansion
\alpha_{i}(v)(1-v^{2})dv%
%TCIMACRO{\tint \limits_{0}^{\infty}}%
%BeginExpansion
{\textstyle\int\limits_{0}^{\infty}}
%EndExpansion
t\exp[-\mathrm{i}(t+\xi\frac{t^{3}}{3})]dt;\label{15}\\
\alpha_{2}  &  =3+v^{2},\ \ \alpha_{3}=2(3-v^{2}),\ \ \sqrt{\xi}=\frac
{\kappa(1-v^{2})}{4},\ \ \kappa^{2}=4r\mu^{2}=-\frac{(Fk)^{2}}{m^{2}H_{0}^{2}%
}.\nonumber
\end{align}
The entering in Eq. (\ref{15}) integrals over $t$ are expressed by derivations
of Airy (the imaginary part) and Hardy (the real part) integrals. Because of
the application conditions, this energy region is overlapped with the
considered above. At $\kappa\ll1,$ we have for the integrals entering into the
real part of $\kappa_{i}$%

\begin{equation}%
%TCIMACRO{\tint \limits_{0}^{\infty}}%
%BeginExpansion
{\textstyle\int\limits_{0}^{\infty}}
%EndExpansion
t\cos tdt=-1,\ \ \
%TCIMACRO{\tint \limits_{0}^{1}}%
%BeginExpansion
{\textstyle\int\limits_{0}^{1}}
%EndExpansion
\alpha_{2}(1-v^{2})dv=\frac{32}{15},\ \ \ \
%TCIMACRO{\tint \limits_{0}^{1}}%
%BeginExpansion
{\textstyle\int\limits_{0}^{1}}
%EndExpansion
\alpha_{3}(1-v^{2})dv=\frac{56}{15}. \label{16}%
\end{equation}
These expression coincides with (\ref{8}).

At calculation of the imaginary part of the integral over $t$ in Eq.
(\ref{15}), we extend the integration to the left real semiaxis because of the
integrand parity. After that, the stationary phase method can be used (
$t_{0}=-$ $\mathrm{i}$ $/\sqrt{\xi}$ ). As a result of the standard procedure
of above method, we have%

\begin{align}
\frac{1}{2}%
%TCIMACRO{\tint \limits_{-\infty}^{\infty}}%
%BeginExpansion
{\textstyle\int\limits_{-\infty}^{\infty}}
%EndExpansion
t\exp[-\mathrm{i}(t+\xi\frac{t^{3}}{3})]dt  &  =\frac{t_{0}}{2}\sqrt{\frac
{\pi}{\mathrm{i}t_{0}\xi}}\exp[-\mathrm{i}(t_{0}+\xi\frac{t_{0}^{3}}%
{3})]=\label{17}\\
-\mathrm{i}\frac{\sqrt{\pi}}{2}\xi^{-3/4}\exp\left(  -\frac{2}{3\sqrt{\xi}%
}\right)   &  =-4\mathrm{i}\sqrt{\pi}\kappa^{-3/2}\left(  1-v^{2}\right)
^{-3/2}\exp\left(  -\frac{8}{3\kappa(1-v^{2})}\right)  .\nonumber
\end{align}
Substituting the obtained expression into Eq. (\ref{15}) and fulfilling the
integration over $v,$ keeping in mind the small $v$ contributes, we get:%

\begin{equation}
\kappa_{2}=-\mathrm{i}\sqrt{\frac{3}{32}}\alpha m^{2}\kappa\exp\left(
-\frac{8}{3\kappa}\right)  ,\ \ \ \kappa_{3}=2\kappa_{2}. \label{18}%
\end{equation}
The condition of validity SQA at $\kappa\ll1,\ r\gg1,$ we can receive
expanding Eq. (\ref{014}) over $r.$This condition has a form $r^{3}\gg\mu
^{-2}.$

At $\kappa\gg1$ ($\xi\gg1$), the small $t$ contributes to the integral
(\ref{15}) ( $\xi t^{3}\sim1),$ and the linear over $t$ term can be omit in
the argument of exponent Eq. (\ref{15}) . Carrying out the change of variable :%

\begin{equation}
\xi t^{3}/3=-\mathrm{i}x,\ \ \ t=\exp\left(  \frac{-\mathrm{i}\pi}{6}\right)
\left(  \frac{3x}{\xi}\right)  ^{1/3}, \label{19}%
\end{equation}
one obtains:%

\begin{equation}
\kappa_{i}=\frac{\alpha m^{2}\kappa^{2}}{24\pi}\exp\left(  \frac
{-\mathrm{i}\pi}{3}\right)  \frac{1}{3}\left(  \frac{48}{\kappa^{2}}\right)
^{2/3}\Gamma\left(  \frac{2}{3}\right)
%TCIMACRO{\tint \limits_{0}^{1}}%
%BeginExpansion
{\textstyle\int\limits_{0}^{1}}
%EndExpansion
dv\alpha_{i}(v)(1-v^{2})^{-1/3}. \label{20}%
\end{equation}
After integration over $v,$ we have:%

\begin{equation}
\kappa_{2}=\frac{\alpha m^{2}(3\kappa)^{2/3}}{7\pi}\frac{\Gamma^{3}\left(
\frac{2}{3}\right)  }{\Gamma\left(  \frac{1}{3}\right)  }(1-\mathrm{i}\sqrt
{3})\simeq\alpha m^{2}\kappa^{2/3},\ \ \ \kappa_{3}=\frac{3}{2}\kappa_{2}.
\label{21}%
\end{equation}

It should be noted that the last expressions for $\kappa_{i}$ do not depend on
the electron mass. For all that, the radiation correction to the mass of
created electron and positron is the same order of values $\Delta m^{2}%
\sim\alpha m^{2}\kappa^{2/3}.$ Because of this, the present consideration is
valid formally at the photon energy $\omega$ when the condition $\alpha
\kappa^{2/3}\ll1$ is fulfilled.

\bigskip The work was supported by the Ministry of Education and Science of
the Russian Federation.

\end{document}